\documentclass[acmsmall]{acmart}

\usepackage{xcolor}
\usepackage{amsmath}
\usepackage{algorithm, algorithmic}
\usepackage{graphicx}
\usepackage{qcircuit}
\usepackage{booktabs}
\usepackage{dcolumn}
\usepackage{bm}

\newcommand{\Tr}{\mathrm{Tr}}
\newcommand{\bytes}{\text{ Bytes}}

\acmJournal{TQC}
\acmVolume{X}
\acmNumber{X}
\acmArticle{X}
\acmYear{2026}

\setcopyright{acmcopyright}
\acmDOI{10.1145/XXXXXXX.XXXXXXX}

\begin{document}
\numberwithin{equation}{section}

\title{Cutting Quantum Circuits Beyond Qubits}

\author{Manav Seksaria}
\affiliation{
  \institution{Indian Institute of Technology Madras}
  \department{Department of Electrical Engineering}
  \city{Chennai}
  \country{India}
}

\author{Anil Prabhakar}
\affiliation{
  \institution{Indian Institute of Technology Madras}
  \department{Department of Electrical Engineering}
  \city{Chennai}
  \country{India}
}

\date{\today}

\begin{abstract}
We extend quantum circuit cutting to heterogeneous registers comprising mixed-dimensional qudits. By decomposing non-local interactions into tensor products of local generalised Gell-Mann matrices, we enable the simulation and execution of high-dimensional circuits on disconnected hardware fragments. We validate this framework on qubit--qutrit ($2$--$3$) interfaces, achieving exact state reconstruction with a Total Variation Distance of 0 within single-precision floating-point tolerance. Furthermore, we demonstrate the memory advantage in an 8-particle, dimension-8 system, reducing memory usage from 128 MB to 64 KB per circuit.
\end{abstract}

\maketitle

\section{Introduction}
The scalability of quantum computers remains one of the most significant hurdles in the Noisy Intermediate-Scale Quantum (NISQ) era~\cite{Preskill2018}. Low qubit counts, restricted connectivity, and short coherence times limit current hardware. To transcend these physical limitations without waiting for hardware breakthroughs, distinct algorithmic approaches have emerged. Foremost among these is Distributed Quantum Computing (DQC)~\cite{Monroe2014, VanMeter2016}, which seeks to aggregate the computational power of multiple small quantum processors.

A key enabler of DQC is ``circuit cutting'', which allows a large quantum circuit to be partitioned into smaller subcircuits to be executed on separate, smaller quantum registers. The non-local dependencies between the partitions are severed and replaced by a sequence of local measurement and preparation operations~\cite{Bravyi2016, Peng2020}, which are then classically combined.

There is also a growing interest in moving beyond the two-level qubit paradigm. Many physical platforms naturally possess more than two accessible energy levels~\cite{Blok2021, Ringbauer2022, Wang2020}. By utilising these higher-dimensional states, or qudits, one can access a larger Hilbert space with fewer physical carriers. However, existing circuit cutting frameworks are predominantly qubit-centric. They rely heavily on the Pauli operator basis ($I, X, Y, Z$) to decompose gates into tensor products of local operations~\cite{Mitarai2021}, leaving a gap in methodology for heterogeneous architectures.

In this work, we extend the formalism of circuit cutting to heterogeneous quantum registers comprising mixed-dimensional qudits. We address the challenge of decomposing non-local interactions between particles of dimensions $d_1$ and $d_2$ where $d_1 \neq d_2$.

\section{Primitive Search}\label{sec:primitive}
\subsection{Decomposition}\label{sec:decomposition}
As an example, from~\cite{Tang2021}, we decompose the qubit CX gate into multiple single-qubit gates:
\begin{equation}
\text{CX} = \frac{1}{2}(I \otimes I + Z \otimes I + I \otimes X - Z \otimes X).
\end{equation}
The decomposition would amount to applying the gates $II$, $ZI$, $IX$ and $-ZX$ on the two qubits respectively, in place of the CX, across four runs of the circuit. We then reconstruct the output state by linearly combining the outputs from the four runs, each weighted by the corresponding decomposition coefficient.

The next step is to generalise the CX and then decompose it into higher dimensions. We define CX from first principles as:
\begin{equation}
 \text{CX}_{q_1, q_2} = I \otimes |0\rangle\langle 0| + X \otimes |1\rangle\langle 1|.
\end{equation}
Similarly, the generalised CX gate, often referred to as the CSUM gate, for qudits of dimension $d$ is defined as:
\begin{equation}
 \text{CX}_{d} = \sum_{j=0}^{d-1} X^j \otimes |j\rangle\langle j|, \text{ where }
 X = \sum_{j=0}^{d-1} |(j+1) \pmod{d}\rangle\langle j|.
\end{equation}
In higher dimensions, we use the generalised Gell-Mann matrices as a basis for qudits, with the Pauli gates forming a special case for $d=2$~\cite{Wang2020}. The Gell-Mann matrices form an orthogonal basis for operators acting on a $d$-dimensional Hilbert space. The Gell-Mann gates are indexed by $j, k$ and $l$, where $1 \leq j < k \leq d$ and $1 \leq l \leq d-1$. Evidently, only the Gell-Mann matrices are required to decompose the gates for our use case.

Using the Gell-Mann matrices indexed by $k$, in dimension $d$, we begin by defining our bases in each dimension $(d_1, d_2)$:
\begin{align}
\mathcal{B}_1 &= \{I\} \cup \{G_{k}^{(d_1)} : 1 \leq k < d_1^2 \},\\
\mathcal{B}_2 &= \{I\} \cup \{G_{k}^{(d_2)} : 1 \leq k < d_2^2 \}.
\end{align}
These matrices allow us to construct appropriate projectors:
\begin{align}
 P_r = \frac{I}{d_1} + \frac{1}{2} \sum_{k=1}^{d_1^2-1} \langle r | G_k^{(d_1)} | r \rangle G_k^{(d_1)}.
\end{align}
We recall the definitions of the generalised $X, CX$ gates:
\begin{align}
 X^r = \sum_{j=0}^{d_2-1} |j+r\rangle\langle j|,
 \quad
 \text{CX}_{d_1,d_2} = \sum_{r=0}^{d_1-1} P_r \otimes X^r.
\end{align}
Then, for each $r$, we obtain the decomposed gates as:
\begin{align}
P_r &= \sum_{A \in \mathcal{B}_1} a_A^{(r)} A
\text{ where }
a_A^{(r)} = \frac{\Tr(P_r A)}{\Tr(A^2)},\\
X^r &= \sum_{B \in \mathcal{B}_2} b_B^{(r)} B
\text{ where }
b_B^{(r)} = \frac{\Tr(X^r B)}{\Tr(B^2)},
\end{align}
finally giving us:
\begin{align}
 \text{CX}_{d_1,d_2} = \sum c_i\, A_i \otimes B_i.
\end{align}
as a matrix with dimension $(d_1d_2)^2$.

\subsection{Reconstruction}
We create a test circuit with arbitrary dimensions to test the mechanism.
\begin{equation}
\Qcircuit @C=1em @R=1.2em {
\lstick{|0\rangle_{d_1}} &
\gate{H} &
\qw &
\qw &
\gate{R_y(\pi/3)} &
\gate{R_z(\pi/4)} &
\qw \\
\lstick{|0\rangle_{d_1}} &
\gate{H} &
\ctrl{1} &
\qw &
\gate{R_y(\pi/5)} &
\gate{R_z(\pi/6)} &
\qw \\
\lstick{|0\rangle_{d_2}} &
\gate{H} &
\targ &
\qw &
\gate{R_y(\pi/7)} &
\gate{R_z(\pi/8)} &
\qw \\
\lstick{|0\rangle_{d_2}} &
\gate{H} &
\qw &
\qw &
\gate{R_y(\pi/9)} &
\gate{R_z(\pi/10)} &
\qw
}
\end{equation}
The test uses the Total Variation Distance  defined as,
\begin{equation}
 \text{TVD}(P, Q) = \frac{1}{2} \sum_{x} |P(x) - Q(x)|.
\end{equation}
\begin{table}[h]
\centering
\caption{Comparison of original and stitched probabilities for a qubit--qubit system. We can see that we can achieve a TVD of 0.0. While we have rounded to five significant figures here, in practice we can achieve a TVD of 0 with \texttt{fp32} precision.}\label{tab:distri}
\begin{tabular}{lccc}
\hline
State & Original & Stitched & Diff \\
\hline
$\lvert 0000\rangle$ & 0.00129 & 0.00129 & 0.00000 \\
$\lvert 0001\rangle$ & 0.00262 & 0.00262 & 0.00000 \\
$\lvert 0010\rangle$ & 0.00326 & 0.00326 & 0.00000 \\
$\lvert 0011\rangle$ & 0.00664 & 0.00664 & 0.00000 \\
$\lvert 0100\rangle$ & 0.00495 & 0.00495 & 0.00000 \\
$\lvert 0101\rangle$ & 0.01010 & 0.01010 & 0.00000 \\
$\lvert 0110\rangle$ & 0.01254 & 0.01254 & 0.00000 \\
$\lvert 0111\rangle$ & 0.02558 & 0.02558 & 0.00000 \\
$\lvert 1000\rangle$ & 0.01791 & 0.01791 & 0.00000 \\
$\lvert 1001\rangle$ & 0.03652 & 0.03652 & 0.00000 \\
$\lvert 1010\rangle$ & 0.04536 & 0.04536 & 0.00000 \\
$\lvert 1011\rangle$ & 0.09251 & 0.09251 & 0.00000 \\
$\lvert 1100\rangle$ & 0.06898 & 0.06898 & 0.00000 \\
$\lvert 1101\rangle$ & 0.14069 & 0.14069 & 0.00000 \\
$\lvert 1110\rangle$ & 0.17471 & 0.17471 & 0.00000 \\
$\lvert 1111\rangle$ & 0.35634 & 0.35634 & 0.00000 \\
\hline
\end{tabular}
\end{table}
Table \ref{tab:distri} gives the results for the primary test case of a qubit--qubit cut. We also tested the qutrit-qutrit and obtained a TVD of 0.0.

The next step is to generalise the process for a qubit-qutrit cut, which requires reconstruction over asymmetric basis sets and unequal bases. Algorithm \ref{algo:recon} shows the procedure for reconstruction of the probabilities of systems with mixed bases. Depending on the simulation framework used, one may need to re-permute the qubits, as we have had to do, to flip big-endian qudits into little-endian qudits.
\begin{algorithm}[H]
\caption{Reconstruction of Probabilities}\label{algo:recon}
\begin{algorithmic}[1]
\STATE \textbf{Input Variables:}
\STATE \quad $\mathbf{A}$: Stitched amplitude vector
\STATE \quad $\mathbf{b}$: Ordered list of base dims (e.g., $[D_2, D_2, D_1, D_1]$)
\STATE \quad $\pi$: Permutation map between cut and logical indices
\STATE \quad $M = \text{length}(\mathbf{b})$
\STATE \quad $N = \text{length}(\mathbf{A}) = \prod \mathbf{b}$
\STATE \quad $T$: Temporary variable
\STATE \textbf{Output:}
\STATE \quad $P$: Map of logical state strings to probabilities

 \,
\FOR{$k \gets 0$ to $N-1$}
 \STATE \textbf{Mixed-Radix Decomposition}
 \STATE $T \gets k$
 \STATE Let $\mathbf{d}$ be an array of size $M$
 \FOR{$j \in [0, M-1]$}
 \STATE \quad $\mathbf{d}[j] \gets T \pmod{\mathbf{b}[j]}$
 \STATE \quad $T \gets \lfloor T / \mathbf{b}[j] \rfloor$
 \ENDFOR
 \\
 \,
 \STATE \textbf{Permutation to Logical Order}
 \STATE Let $\mathbf{d}'$ be an array of size $M$
 \FOR{$j \in [0, M-1]$}
 \STATE \quad $\mathbf{d}'[j] \gets \mathbf{d}[\pi[j]]$
 \ENDFOR
 \\
 \,
 \STATE \textbf{Probability Assignment}
 \STATE $s \gets \text{Join } \mathbf{d}' \text{ into string}$
 \STATE $P[s] \gets |\mathbf{A}[k]|^2$
\ENDFOR

\RETURN $P$
\end{algorithmic}
\end{algorithm}
\begin{table}[h]
\centering
\caption{We can see that even for an asymmetric system, we can achieve a TVD of 0.}\label{tab:asym}
\begin{tabular}{c|c|c|c}
\hline
State & Original & Stitched & Diff \\
\hline
$\lvert0000\rangle$ & 0.00057 & 0.00057 & 0.00000 \\
$\lvert0001\rangle$ & 0.00117 & 0.00117 & 0.00000 \\
\vdots & \vdots & \vdots & \vdots\\
$\lvert1121\rangle$ & 0.11045 & 0.11045 & 0.00000 \\
$\lvert1122\rangle$ & 0.08230 & 0.08230 & 0.00000 \\
\hline
\end{tabular}
\end{table}
From Table \ref{tab:asym}, we see that we have demonstrated cutting on a two--qubit and a two--qutrit system. This will allow us to run mixed circuits without co-locating qubits and qutrits on the same physical chip, or, in general, to optimise dits of different dimensions individually and place them on different chips/chiplets. While this example uses our dummy circuit from above, we can apply the method to any circuit with heterogeneous qudits and place the cut at any position, not just at the halfway mark.

\subsection{Memory and Scaling}
We will now demonstrate a memory advantage on the problem using an eight-particle system of dimension 8, which is cut into two halves. Each cut will then have four particles, each of dimension 8.

Assuming \texttt{complex64} precision (8 Bytes), we can calculate the memory requirement for simulating an 8-qudit system of dimension eight as:
\begin{align}
\text{Memory}
 = 8^{8} \times 8\bytes = 2^{24} \times 8\bytes 
 = 128 \text{ MB}.
\end{align}
While this is not too large for modern systems, we can use Linux cgroups to limit the process's memory to demonstrate the advantage of circuit cutting. If we cut the circuit into two 4-qudit subcircuits, we can simulate each subcircuit separately and then stitch the results together. The memory requirement for each 4-qudit subcircuit of dimension 8 is:
\begin{align}
\text{Memory}
 = 2 \times (8^4 \times 8\bytes) = 2 \times (2^{12} \times 8\bytes)  = 64 \text{ KB}.
\end{align}
While we save memory, we pay the price in additional circuit evaluations and classical post-processing time. Further, a circuit-cutting evaluation for circuits larger than the memory permits will have to be performed using a disk-assisted cache that periodically writes intermediate states to disk, further increasing the evaluation time.

With a limited memory of 150MB, it took us $\approx130$s to simulate the whole circuit, while the cut circuit took $\approx1350$s to simulate with a TVD of $0.00000$, across 532 subcircuit pairs. When the memory was limited to 100MB, the full circuit simulation failed, while the cut circuit simulation still completed in $\approx1400$s. Since we have used a little more than the exact amount of memory the problem requires, there was negligible time overhead due to swapping, or IO, having loaded the whole problem into memory at once.

In general, from Section \ref{sec:decomposition}, for a cut between qudits of dimensions $d_1$ and $d_2$, we will have a basis of size $d_1^2$ and $d_2^2$ respectively. This means that the total number of terms in the decomposition will be $d_1 \cdot (d_1 d_2)^2$. However, in practice, we rarely get even close to the theoretical maximum, as many decomposition coefficients are zero or negligible. We can set several thresholds for coefficient truncation and check the number of terms retained versus the TVD achieved for our dummy circuit.

\begin{figure}[h]
 \centering
 \includegraphics[width=0.8\linewidth]{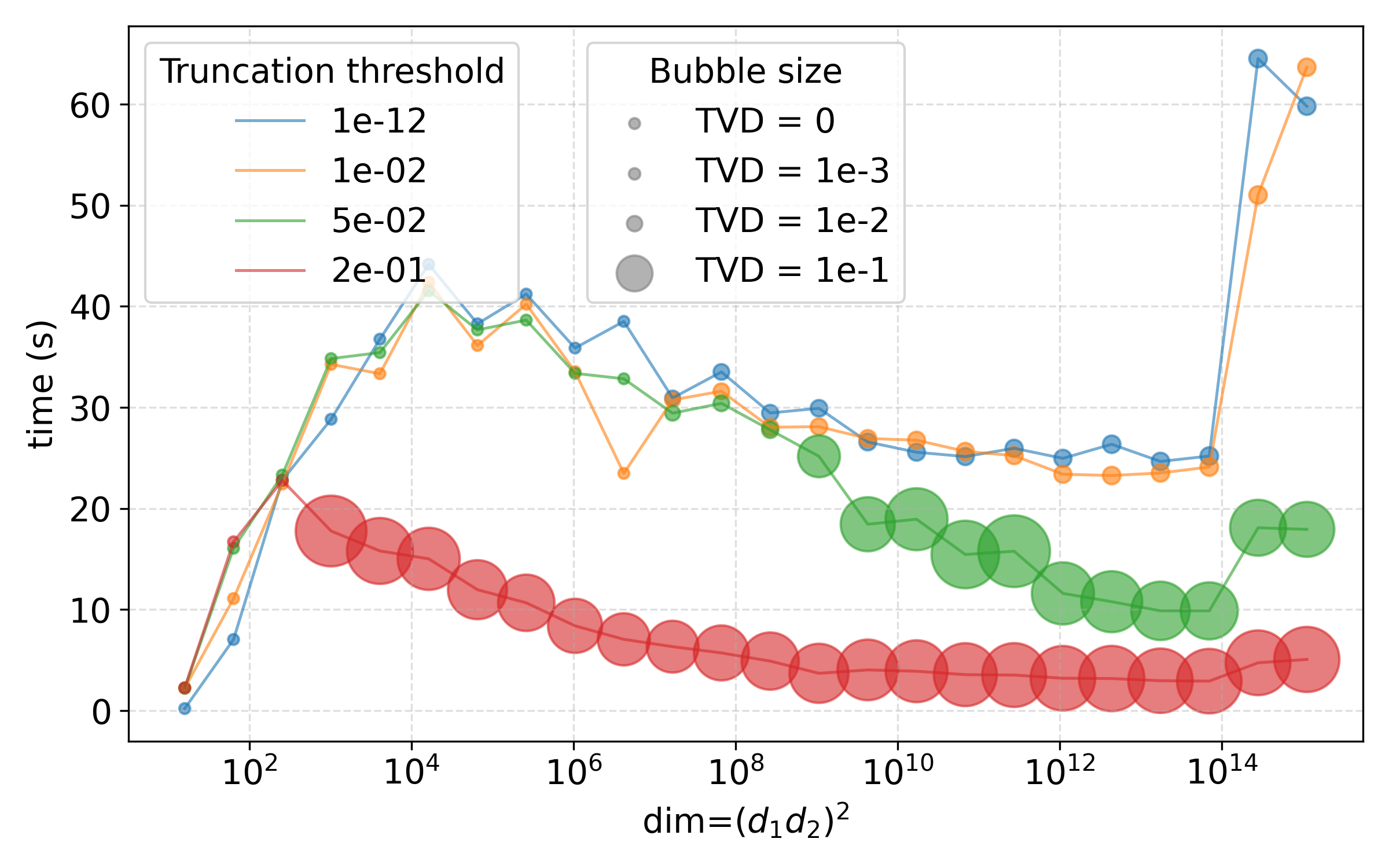}
 \caption{The scaling of TVD and simulation time with increasing system size for different truncation thresholds shows us that even when we truncate coefficients up to $10^{-2}$, while we can maintain a low TVD, there is not much time saved. However, for higher truncation thresholds, we can see a significant reduction in simulation time at the cost of increased TVD.}
 \label{fig:scaling}
\end{figure}

From Fig \ref{fig:scaling}, we can see that even for large systems of dimension $10^6$, a truncation of up to $10^ {-2}$ still gives us a TVD of $0.0$ to at least three decimal places. This trend is also observed even with a truncation of $5\times 10^{-2}$; however, after a system size of $10^9$, the TVD starts to increase. We suspect this increase is due to both the accumulation of truncation errors across multiple terms and to numerical instability arising from floating-point precision limits. Beyond a system size of $10^{14}$, we can see that if we are willing to accept a TVD of $10\%$, then we can finish the computation in less than a third of the time taken for the full computation.

We can also check the scaling of the simulation time with increasing qudit dimensions for different truncation thresholds. As system size increases, we expect truncation to save more time, as the number of negligible terms in the decomposition increases.

\section{Schmidt Decomposition}
\begin{figure}[h]
 \centering
 \includegraphics[width=0.8\linewidth]{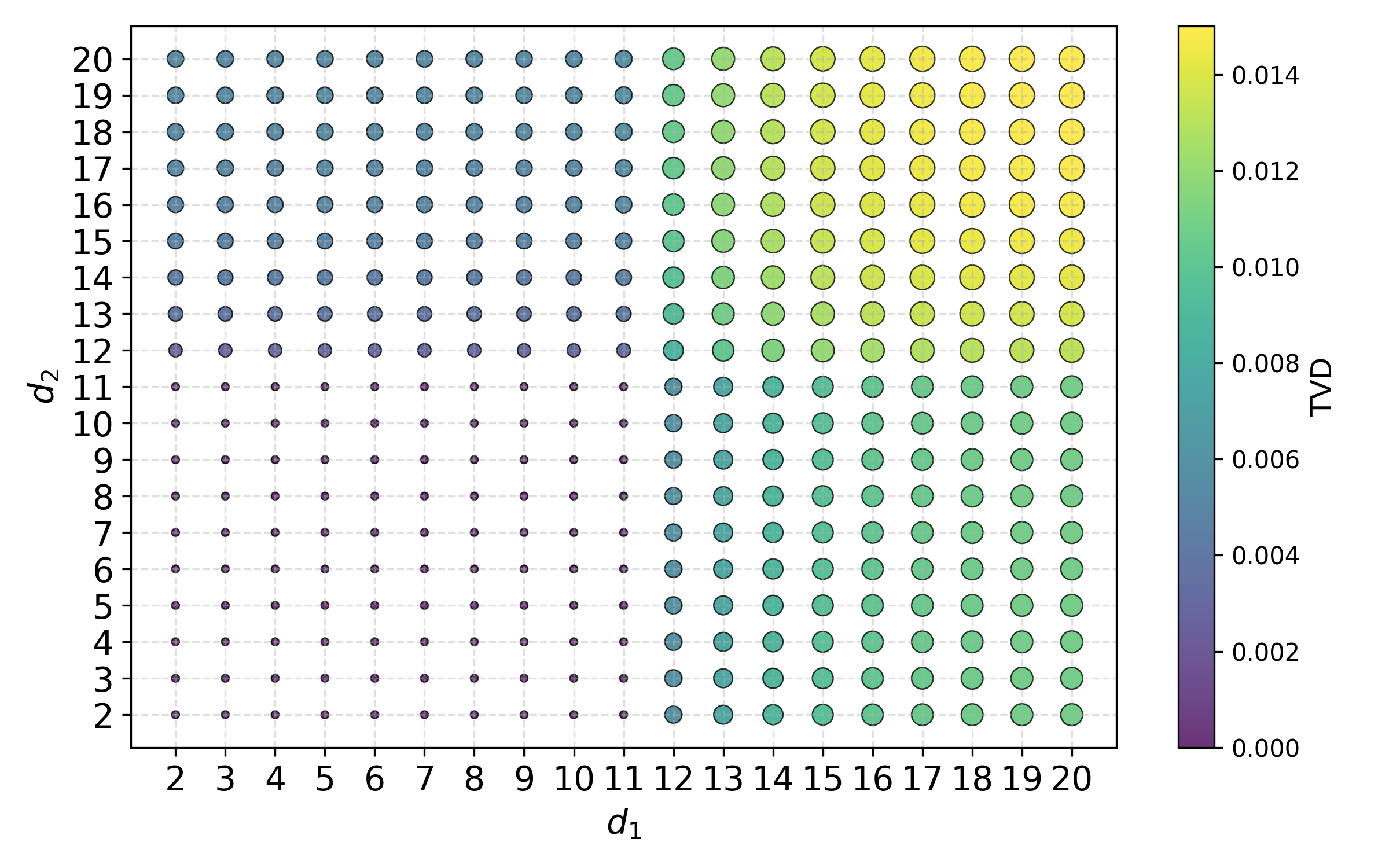}
 \caption{The dependence of TVD on the dimensions $d_1, d_2$ for the CX gate. The error is zero when $d_1,d_2 \leq 11$ , as seen by the dots at TVD $= 0.000$.}
 \label{fig:schmidt-tvd}
\end{figure}

In Section \ref{sec:primitive}, we opted to go for a full basis search. Although a full basis search is computationally expensive, it allows us to leverage hardware-native gates more efficiently, since these are usually implemented as variations of standard basis vectors, such as the Pauli gates in qubits. However, as we move to higher dimensions, the number of bases increases quadratically; therefore, it stands to reason that, for larger dimensions, it may be easier to allow arbitrary unitary gates directly rather than specific basic gates.
Given a unitary $U$ acting on qudits of dimensions $d_1$ and $d_2$, we can perform a singular value decomposition (SVD) to obtain:
\begin{equation}
 U = \sum_{i=1}^{r} \sigma_i\, U_i^{(d_1)} \otimes V_i^{(d_2)},
\end{equation}
where $r=\min(d_1, d_2)$ is the rank of the decomposition, $\sigma_i$ are the singular values, and $U_i^{(d_1)}$, $V_i^{(d_2)}$ are the local unitaries acting on the respective qudits.
Since Schmidt decomposition gives us the minimum number of terms, it also results in a much lower TVD. Fig \ref{fig:schmidt-tvd} shows how TVD depends on $d_1$, in the optimal case. The TVD achieved even for a $12, 12$ cut is $<0.01$ due to there being only 12 terms in the decomposition. This is in contrast to the full basis search, which would have had $225$ terms for the same result.

On a quantum computer, a gate cut does not reduce the depth (and therefore does not reduce the execution time); it does reduce the number of qudits required. On classical hardware, most programs are memory-bound, so reducing the number of qudits directly translates into a reduction in execution time.
Consider the previous example of an 8-qudit system of dimension 8. When running a full basis search with no truncation, we obtain 63 terms in the decomposition, giving us a memory consumption of $64\text{ KB} \times 63 \times 2 \approx 7.9 \text{ MB}$. However, using the Schmidt decomposition, we only require 8 terms in the decomposition, giving us a memory consumption of $64\text{ KB} \times 8 \times 2 = 1 \text{ MB}$. A much lower memory consumption directly translates into lower execution time, as we can fit the entire computation into the CPU cache.

\begin{figure}[h]
 \centering
 \includegraphics[width=0.8\linewidth]{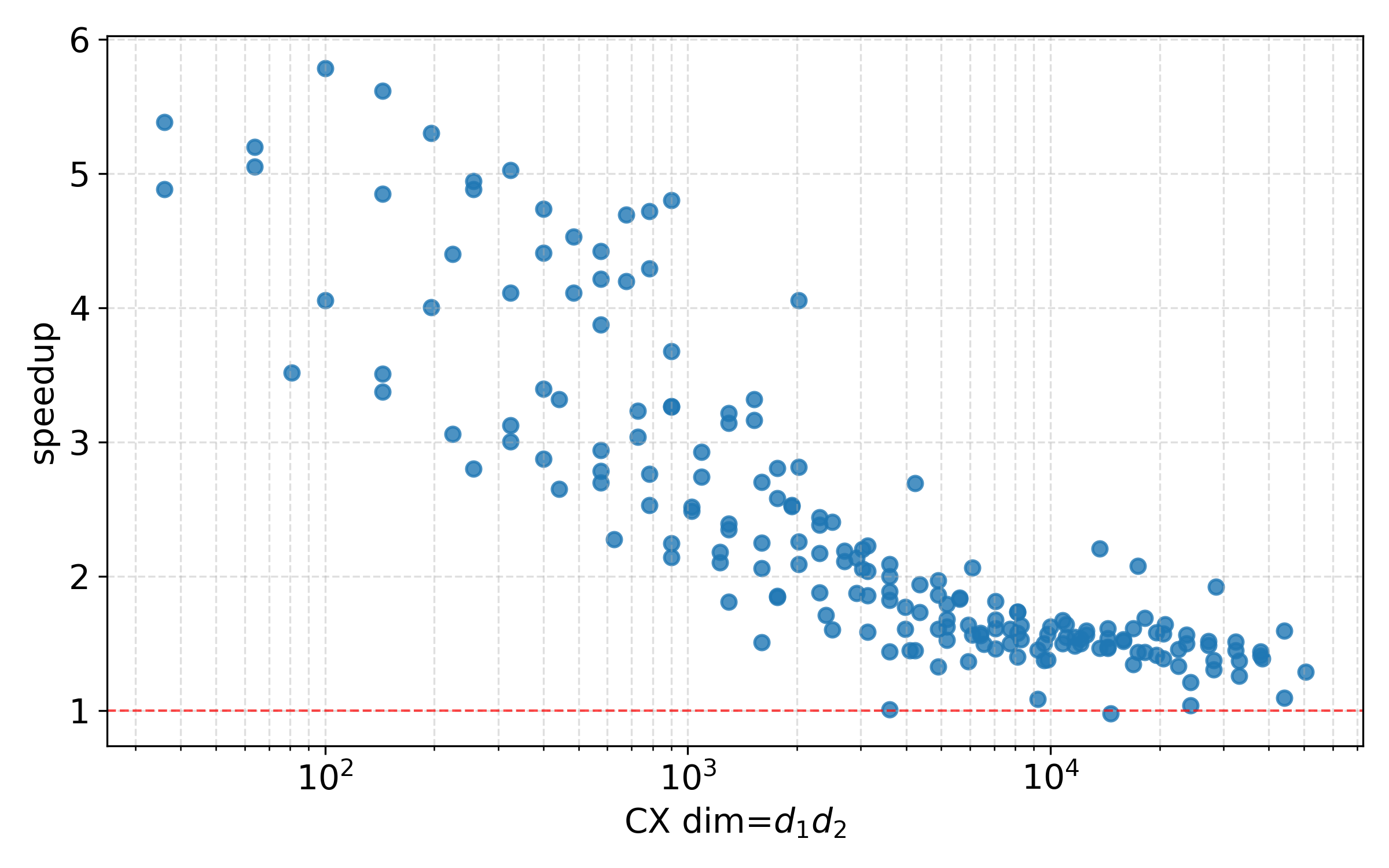}
 \caption{Improvement in execution time using Schmidt decomposition for a 4-qudit system, across multiple runs with different CX dimension, $(d_1 d_2)^2$.}
  \label{fig:schmidt-time}
\end{figure}

We define the multiplication factor, or speedup, as
\begin{equation}
    M = \frac{\text{Time taken by an uncut circuit}}{\text{Time taken by the cut circuits with Schmidt decomposition}}
\end{equation}
and plot the improvement in execution time in Fig. \ref{fig:schmidt-time} as the CX dimension increases. Each point in the plot is the average speedup we observed across 12 runs.

As one would expect, as the dimension increases, the speedup decreases because the stitching overhead increases. For our 4-qudit system, even up to a CX dimension of $50^2$, we see more than a 5x speedup when using the Schmidt decomposition compared to simulating the uncut circuit. As one would expect, the exact point at which we cross over from a speedup to a slowdown varies by both hardware and circuit.

All simulations in this work were run on a 2022 Apple MacBook Air, with the base M2 processor. The backend was provided by \texttt{qudit}]~\cite{Seksaria2025} running on python $3.12.11$.

\section{Conclusion}
In this work, we have generalised the circuit cutting formalism to arbitrary mixed-dimensional quantum systems. By deriving the decomposition of the generalised CX gate using an asymmetric Generalised Gell-Mann basis, we demonstrated that qudits of unequal dimensions—such as qubits and qutrits—can be cut and classically recombined with high fidelity.

Our numerical experiments confirm that the reconstruction is exact, yielding a TVD of $0.00000$ (we have checked it up to the $12^\text{th}$ decimal) for both homogeneous and heterogeneous cuts. The primary advantage of this technique lies in the compression of memory and connectivity. As demonstrated in our dimension-8 stress test, we simulate an 8-qudit system using only 64 KB of memory per subcircuit, whereas the monolithic simulation required 128 MB. A space--time trade-off and allowing heterogeneous particles to be separated both contribute a small step towards large-scale distributed quantum computing.

Future work will focus on generalising this work to allow multiple simultaneous cuts.

\bibliographystyle{ACM-Reference-Format}
\bibliography{ref}

\end{document}